\documentclass[11pt,a4paper]{article}
\usepackage[hyperref]{acl2020}
\usepackage{times}
\usepackage{latexsym}
\usepackage{graphicx,graphics,multirow,color}

\usepackage{rotating}
\usepackage{framed}
\usepackage{todonotes}
\usepackage{amsmath}
\usepackage{subcaption}
\usepackage{lscape}

\usepackage{microtype}

\aclfinalcopy % Uncomment this line for the final submission
\usepackage{xcolor,colortbl}

\definecolor{Gray}{gray}{0.85}
\definecolor{LightCyan}{rgb}{0.88,1,1}

\newcolumntype{a}{>{\columncolor{Gray}}c}
\newcolumntype{b}{>{\columncolor{white}}c}

\title{Exploring Recurrent, Memory and Attention Based Architectures for Scoring Interactional Aspects of Human--Machine Text Dialog}

\author{Vikram Ramanarayanan$^\dagger$, Matthew Mulholland$^\ddagger$ \& Debanjan Ghosh$^\ddagger$\\
  Educational Testing Service R\&D\\
  $^\dagger$90 New Montgomery Street, Suite 1500, San Francisco, CA\\
  $^\ddagger$660 Rosedale Rd., Princeton, NJ \\
  \texttt{vramanarayanan,mmulholland,dghosh@ets.org} \\
 }
\date{}

\begin{document}
\maketitle
\begin{abstract}
An important step towards enabling English language learners to improve their conversational speaking proficiency involves automated scoring of multiple aspects of interactional competence and subsequent targeted feedback. This paper builds on previous work in this direction to investigate multiple neural architectures -- recurrent, attention and memory based -- along with feature-engineered models for the automated scoring of interactional and topic development aspects of text dialog data. We conducted experiments on a conversational database of text dialogs from human learners interacting with a cloud-based dialog system, which were triple-scored along multiple dimensions of conversational proficiency. We find that fusion of multiple architectures performs competently on our automated scoring task relative to expert inter-rater agreements, with (i) hand-engineered features passed to a support vector learner and (ii) transformer-based architectures contributing most prominently to the fusion. 
\end{abstract}

\section{Automated Scoring of Text Dialog}
\label{sec:intro}

%	I. Intro
%	- Intro to (language) learning and assessment and why conversations and dialog are crucial to test the constructs. Current state of the art. Text and speech. Monolog. Interpretability.
% - Dialog prior art. Challenges over monolog. What has been done in earlier work by us?
%	- What are we proposing here? Why? Aims of this paper (how does it improve over previous SIGDIAL work?). 
%	- Overview of the paper
There is an increasing demand for dialog-based learning and assessment solutions at scale, given the rapidly growing language learning and online education marketplace.  %Conversational proficiency is a crucial skill for success in today's workplace \cite{weldy97,oliveri19}. 
Dialog system technologies are one solution capable of addressing and automating this demand at scale \cite{ramanarayanan16a}. However, such conversational technologies need to be able to provide useful and actionable feedback to users in order for them to be widely adopted. Automated scoring of multiple aspects of conversational proficiency is one way to address this need. 
%	- Current state of the art. Text and speech. Monolog. Interpretability -- this is the KEY issue. That's why neural networks can be shut out. For now.
While the automated scoring of text and speech data has been a well-explored topic for several years, particularly for essays and short constructed responses in the case of the former \cite{shermis13,burrows15,madnani17} and monolog speech for the latter \cite{neumeyer00,witt00,xi12,bhat15}), research on the \textit{interpretable} automated scoring of dialog has only recently started gaining traction \cite{evaniniEtAl2015content,LITMAN16,ramanarayanan17a}. 
% Other work has analyzed this scoring problem at the level of each response in the dialog (i.e., each turn) instead of the entire conversation and across multiple dimensions of speaking proficiency \cite{}. %\cite{ramanarayanan17anonymized}.  %The fine-grained proficiency ratings thus enable the possibility of providing more targeted automated feedback about the learner's English proficiency. 
Further, certain dialog constructs such as those pertaining to interaction --  engagement, turn-taking and repair -- are a lot less well-studied as compared to others like delivery and language use. % these specifics unique compared to the monologic speech and it is .  
%One of the challenges with dialog is that there are certain aspects that are unique to it relative to monologic speech, such as those specific to interaction -- like engagement, turn-taking and repair, for instance. 
\newcite{ramanarayanan2019scoring} recently performed a comprehensive examination of the automated scoring of \textit{content} of whole dialog responses based primarily on text features, based on a comprehensive multidimensional rubric and scoring paradigm designed specifically focusing on aspects of interaction. 

This paper aims to expand on the analysis presented in \newcite{ramanarayanan2019scoring} more comprehensively along two directions. First, we also investigate constructs of text dialog scoring rubric pertaining to topic development along with those pertaining to interaction, aiming to understand, for the first time, how various feature-engineering and model-engineering methods perform on a broader range of scoring dimensions. Second, we propose a more comprehensive experimental setup that explores multiple feature-engineered models (that include novel lexical features from the politeness detection literature \cite{danescu2013computational}) and deep learning network architectures -- recurrent, attention and memory based -- for automated scoring. We specifically study LSTM (Long Short-Term Memory) networks with context attention \cite{yang2016hierarchical}, memory networks \cite{weston2014memory,sukhbaatar15}, and the BERT (Bidirectional Encoder Representations from Transformers) family of models \cite{devlin2018bert}. 
% We also introduce novel lexical features to the previous state-of-the-art feature-engineered models that are inspired by prior computational work on politeness detection \cite{danescu2013computational} to better capture certain aspects of topic development and pragmatic appropriateness.
% Finally, we examine the efficacy of fusing multiple architectures at the score-level to improve the automated scoring performance even further. 
Finally, we report performance improvements using score-level fusion of multiple models. 

%First, we aim to investigate multiple neural architectures -- recurrent, attention and memory based -- for automated scoring. 

%\textcolor{red}{Add in a paragraph talking about the state of the art in recurrent, memory and attention architectures.}

\begin{table*}[t]
% \vspace{-5mm}
\caption{\label{tab:scoringRubric} {\it Human scoring rubric for interaction aspects of conversational proficiency. Scores are assigned on a Likert scale from 1-4 ranging from low to high proficiency. A score of 0 is assigned when there were issues with audio quality or system malfunction or off-topic or empty responses.}}
\centerline{
\scalebox{0.9}{
\begin{tabular}{ | l | c | p{11cm} | }
\hline
Construct	& Sub-construct & Description  \\
\hline
%\multirow{4}{*}{Linguistic Control} &  Fluency	& 	Examines to what extent the response includes pauses at appropriate locations to formulate ideas and good tempo with minimal hesitation. \\
%\cline{2-3}
%& Pronunciation	&  Examines to what extent the response L1 influence and word-level pronunciation impacts intelligibility.  \\
%\cline{2-3}
%& Rhythm 	& Examines the extent to which appropriate sentence-level intonation and stress is used to convey meaning without hindering intelligibility.  \\
%\cline{2-3}
%& Grammar \& Vocabulary	& Examines the extent to which range of grammar structures and vocabulary is accurately used to express clear and precise meanings.   \\
%\hline
\multirow{4}{*}{Topic Development} &  Topic	& 	Examines to what extent the responses are uniformly on topic and relevant. \\
\cline{2-3}
& Elaboration	&  Examines the extent to which arguments are developed taking into account dialog history and with minimal or no repetition.  \\
\cline{2-3}
& Structure 	& Evaluates the structure of the discourse and chain of reasoning, along with the appropriate use of discourse markers.  \\
\cline{2-3}
& Task	& Evaluates how well the user accomplished the task over the course of the interaction.   \\
\hline
\multirow{4}{*}{Interaction} &  Engagement	& 	Examines the extent to which the user engages with the dialog agent and responds in a thoughtful manner. \\
\cline{2-3}
& Turn Taking	&  Examines the extent to which the user takes the floor at appropriate points in the conversation without noticeable interruptions or gaps.  \\
\cline{2-3}
& Repair 	& Examines the extent to which the user successfully initiates and completes a repair in case of a misunderstanding or error by the dialog agent.  \\
\cline{2-3}
& Appropriateness	& Examines the extent to which the user reacts to the dialog agent in a pragmatically appropriate manner.   \\
\hline
\hline
\multicolumn{2}{|c|}{Overall Holistic Performance} & Measures the overall performance.\\
\hline
\end{tabular}
}
}
%\vspace{-3mm}
\end{table*}

% - Prior art in short response scoring with text features. 

%	- Proposed work. Aims of this paper. 
%	- Overview of the paper

\section{Data}
\label{sec:data}

% \subsection{The HALEF dialog ecosystem}
% \label{sec:HALEF}
% 
%    \begin{figure}[thb]
%   \centering
%   \hspace{-5mm}
%   \includegraphics[scale=.37]{Figs/HALEF_v5_multimodal_crowdsourcing}
%   \caption{The HALEF dialog system used in a crowdsourcing-based iterative bootstapping setup for rapid development and data collection.}
%   \label{fig:HALEF}       % Give a unique label
%   \end{figure}

\subsection{Collection}
\label{sec:AMT}

We analyze a corpus of 2288 conversations of non-native speakers introduced in  \newcite{ramanarayanan2019scoring}. Here, speakers interact with a dialog application designed to test general English speaking competence in workplace scenarios particularly focusing on pragmatic skills. The application %, dubbed ``Request Boss''
 requires participants to interact with their boss and request her for a meeting to review presentation slides using pragmatically appropriate language \cite[see][for more details]{timpe2017anonymized}. %Details of the dialog system used for data collection are described in  \newcite{ramanarayanan2019scoring}. 
  To develop and deploy this application, we leveraged %HALEF\footnote{\texttt{http://halef.org}}, 
 an open-source modular cloud-based dialog system that is compatible with multiple W3C and open industry standards \cite{ramanarayanan16anonymized}. 
%\cite{ramanarayanan16}

 \iffalse
 To develop and deploy this application, we leveraged %HALEF\footnote{\texttt{http://halef.org}}, 
 an open-source modular cloud-based dialog system that is compatible with multiple W3C and open industry standards \cite{ramanarayanan16anonymized}. 
%\cite{ramanarayanan16}
The %HALEF 
dialog system logs speech data collected from participants to a data warehouse, which is then transcribed and scored.  
\fi

% \begin{table*}[tbh]
%  %\vspace{-5mm}
% \caption{\label{tab:features} {\it c-rater ML features used for machine scoring.}}
% \centerline{
% \scalebox{0.9}{
% \begin{tabular}{  c | p{10cm}  }
% \hline
% Feature & Description  \\
% \hline
% Word \textit{n}-grams	& Word \textit{n}-­grams are collected for n = 1 to 2. This feature captures patterns about vocabulary usage (key words) in responses. \\
% \hline
% Character \textit{n}-grams &	Character \textit{n}­-grams (including punctuation and whitespace) are collected for \textit{n} = 2 to 5. This feature captures patterns that abstract away from spelling errors and other grammatical errors. \\
% \hline
% Response length &	The number of characters in a response (defined as \textit{log(chars)}, where \textit{chars} represents the total number of characters in a response). \\
% \hline
% Syntactic dependencies & A feature that captures grammatical relationships between individual words in a sentence. This feature captures linguistic information about ``who did what to whom'' and abstracts away from a simple unordered set of key words. \\
% \hline
% \end{tabular}
% }
% }
% %\vspace{-3mm}
% \end{table*}

\subsection{Human Scoring}
\label{sec:scoring}

Each of the 2288 dialog responses were triple scored by human expert raters on a custom-designed rubric. The rubric defined 12 sub-constructs under the 3 broad constructs of linguistic control, topic development and interaction, apart from an overall holistic score. This study investigates the topic development construct for the first time in addition to interaction. %. Although \newcite{ramanarayanan2019scoring} conducted experiments on the interaction aspect, in this study, for the first time, we are addressing the topic development constructs as well. 
See Table \ref{tab:scoringRubric} for specific details of the constructs examined.  

% \DG{DG: explain the differences between score 1 - 4}
 
% However, for purposes of this first study, we will focus on the relatively understudied interaction construct, in particular aspects of engagement, turn-taking, repair and (pragmatic) appropriateness. See Table \ref{tab:scoringRubric} for more details. The expert raters were scoring leaders with significant experience in scoring various spoken and written assessments of English language proficiency. We used an automatic randomized design to assign three (out of eight possible) raters to every full-call response such that (i) all raters had a commensurate number of responses to rate, and (ii) the same group of raters did not rate the same set of files (achieved by randomization; this prevents unwitting biases due to individual rater profiles creeping into the overall score analysis).  

%\input{features}
\section{Automated Scoring Methods} 
\label{sec:scoring_expts}

We first describe the hand-engineered feature set used in conjunction with a linear support vector machine (SVM) classifier. We then describe, in turn, the recurrent, memory and attention based architectures investigated in this paper. 
%We then analyze human performance (by examining inter-rater statistics) and use this to benchmark the performance of machine scoring methods. 
We trained all automated scoring models in this paper to predict valid dialog-level scores from 1-4 (we only consider dialogs with a non-zero score to train scoring models \footnote{Standard practice in automated scoring typically involves training a separate \textit{filtering model} to filter out ``unscorable'' responses, which include responses with no, garbled or out-of-topic audio data. See \citet{higgins11} for a more detailed motivation and rationale for this approach.}). An exception to this is in the case of the memory network, where we predict scores at the turn-level, and then report the dialog level score as the median score across all turns of that dialog. We report the mean performance of scoring systems on a 10-fold cross-validation (CV) experimental setup. 
%We used a cross entropy (log-loss) objective function to optimize learner performance, and fine-tuned hyperparameters such as the regularization coefficient using a grid search method. 
Finally, we report both accuracy and quadratic weighted kappa (which takes into account the ordered nature of the categorical labels) as metrics. %, reported in Table \ref{tab:stats}. 

\subsection{Feature Engineering Approaches}  
\label{sec:skll}

\begin{table*}[tbh]
 \vspace{-4mm}
\caption{\label{tab:features} {\it Content and grammatical structure features used for machine scoring.}}
\centerline{
\scalebox{0.9}{
\begin{tabular}{  c | p{10cm}  }
\hline
Feature & Description  \\
\hline
Word \textit{n}-grams	& Word \textit{n}-grams are collected for n = 1 to 2. This feature captures patterns about vocabulary usage (key words) in responses. \\
\hline
Character \textit{n}-grams &	Character \textit{n}-grams (including whitespace) are collected for \textit{n} = 2 to 5. This feature captures patterns that abstract away from grammatical and other language use errors. \\
\hline
Response length &	Defined as \textit{log(chars)}, where \textit{chars} represents the total number of characters in a response. \\
\hline
Syntactic dependencies & A feature that captures grammatical relationships between individual words in a sentence. This feature captures linguistic information about ``who did what to whom'' and abstracts away from a simple unordered set of key words. \\
\hline
Discourse strategy & Features based on presence or absence of specific words in the response that represent different discourse strategies (see Table \ref{table:politeexamples} for examples of politeness strategies). \\
\hline
\end{tabular}
}
}
%\vspace{-1mm}
\end{table*}

We examine two sets of features here.  First, we re-implement the features from \newcite{ramanarayanan2019scoring}, i.e., features that explicitly capture content (e.g., word n-grams, character n-grams) and grammatical structures (e.g., dependency trees). We briefly summarize them in Table \ref{tab:features}. We have found them to be effective in predicting sub-constructs such as \emph{engagement} and \emph{turn taking} in earlier work.  Second, we introduce nuanced features that are related to the power dynamics of social interactions and are often indicators of whether an interaction went well or not. We hypothesize that features that capture interaction strategies such as gratitude expression or greetings will be particularly useful, given that the corpus involves conversations between a participants and their boss. We therefore focus on features that capture politeness and acknowledgment. These are inspired by \newcite{danescu2013computational}, who conducted a very thorough analysis of politeness strategies employed by Wikipedia and Stack Exchange users. 
Our features capture strategies such as counterfactual modals (``could/would you \dots"), the indicative modal (``can/will you \dots''), deferential back-shift (``I was wondering \dots''), gratitude (``Thank you \dots''), apologies (``I apologize'', ``forgive me''), appreciation, especially at the end of the conversation (``sounds good'', ``works great''), requests (``please review \dots''), greetings (``Hi, hello miss''), mainly in the beginning of the conversation to build a positive relationship, and hedging (``I suggest \dots''). \footnote{We used a vocabulary of hedge words from \texttt{https://github.com/sudhof/politeness.}} These features are binary, indicating, whether a dialog consists a specific politeness strategy. Table \ref{table:politeexamples} presents exemplars of politeness strategies observed in our training corpus. 

\begin{table}
\caption{{\it Politeness strategy exemplars reproduced from the training corpus.}}
\centering
\scalebox{0.9}{
%\begin{tabular}{ |l|p{10cm}| } 
\begin{tabular}{ p{2.1cm}|p{5.4cm} }
\hline
%\multicolumn{1}{|c|}{Platform} & {c}{Turn Type} & 
%\multicolumn{1}{c|}{Turn pairs} \\
  Strategy  &  Example \\
\hline
Counterfactual & \emph{Could} you also review my slides?\\
Indicative & \dots if we \emph{can} meet \dots \\
Deferential & I was \emph{wondering} do you have time\\
Gratitude & I greatly \emph{appreciate} your time.\\
Apology & \emph{Sorry} to bother you \dots \\
Appreciation  & \emph{Sounds good}. I will see you \dots \\
Request & \emph{Please} review the presentation. \\
Greetings & \emph{Hi Hello} Miss Lisa it is good \dots \\
Hedges & \dots and \emph{suggest} me anything \dots \\
\hline
\end{tabular}
}
\label{table:politeexamples}
\end{table}

We used SKLL, an open-source Python package that wraps around the \textit{scikit-learn} package \cite{scikit-learn} to perform machine learning experiments. % to predict valid scores from 1-4 (we removed dialog files that were assigned a score of 0). 
We report the mean performance of linear support vector machines (SVM) where we used a cross entropy (log-loss) objective function to optimize learner performance, and fine-tuned hyperparameters such as the regularization coefficient using a grid search method. %We computed both accuracy and quadratic weighted kappa (which takes into account the ordered nature of the categorical labels) as metrics, reported in Table \ref{tab:stats}. 

%We further tuned and optimized the free parameters of each learner c.  
%and multi-layer perceptron regressors. The former allows us to interpret how the algorithm performs while the latter is used for comparison purposes to understand how deep neural networks might perform on this task given the data we have. In our case, we found that the SVM classifier beat the MLP across the board, possibly because our feature space is sparse and high-dimensional, consisting of binary presence/absence features. We ran

\begin{figure}[htb]
\centering
\begin{framed}
\includegraphics[width=7.5cm]{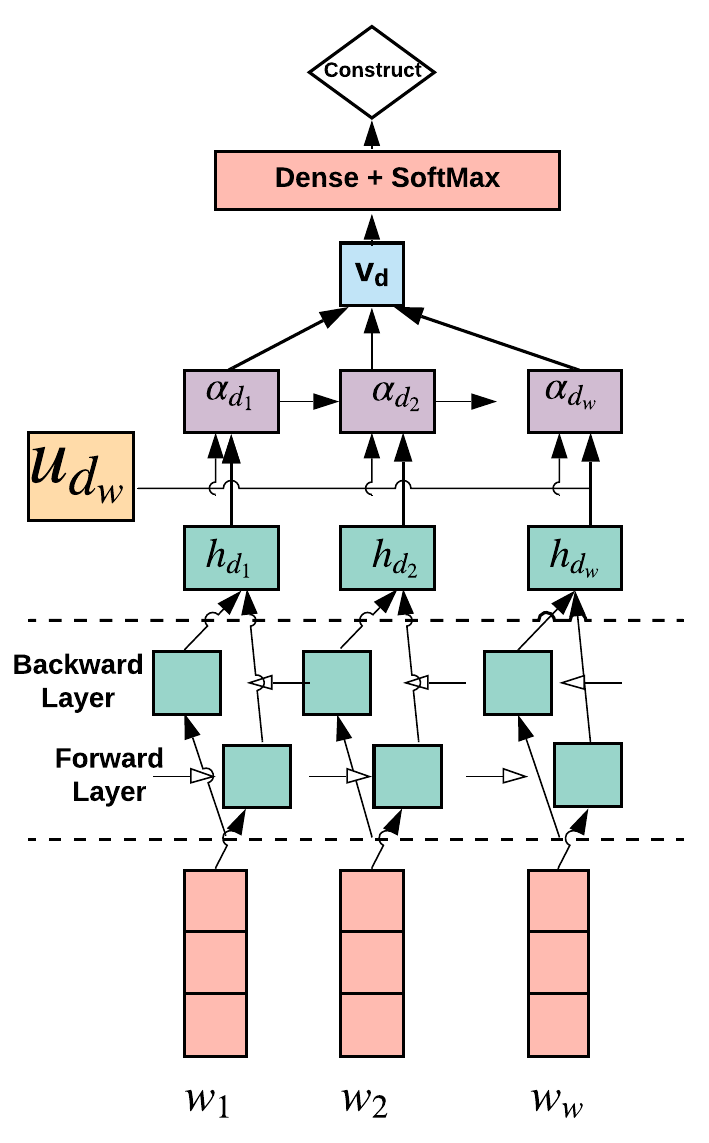}
\end{framed}
\caption{Word-level BiLSTM + Attention Network for dialog}
\label{figure:bilstm}
\end{figure}

\begin{figure*}[hbt]
% \vspace{-5mm}
\centering
\includegraphics[width=170mm]{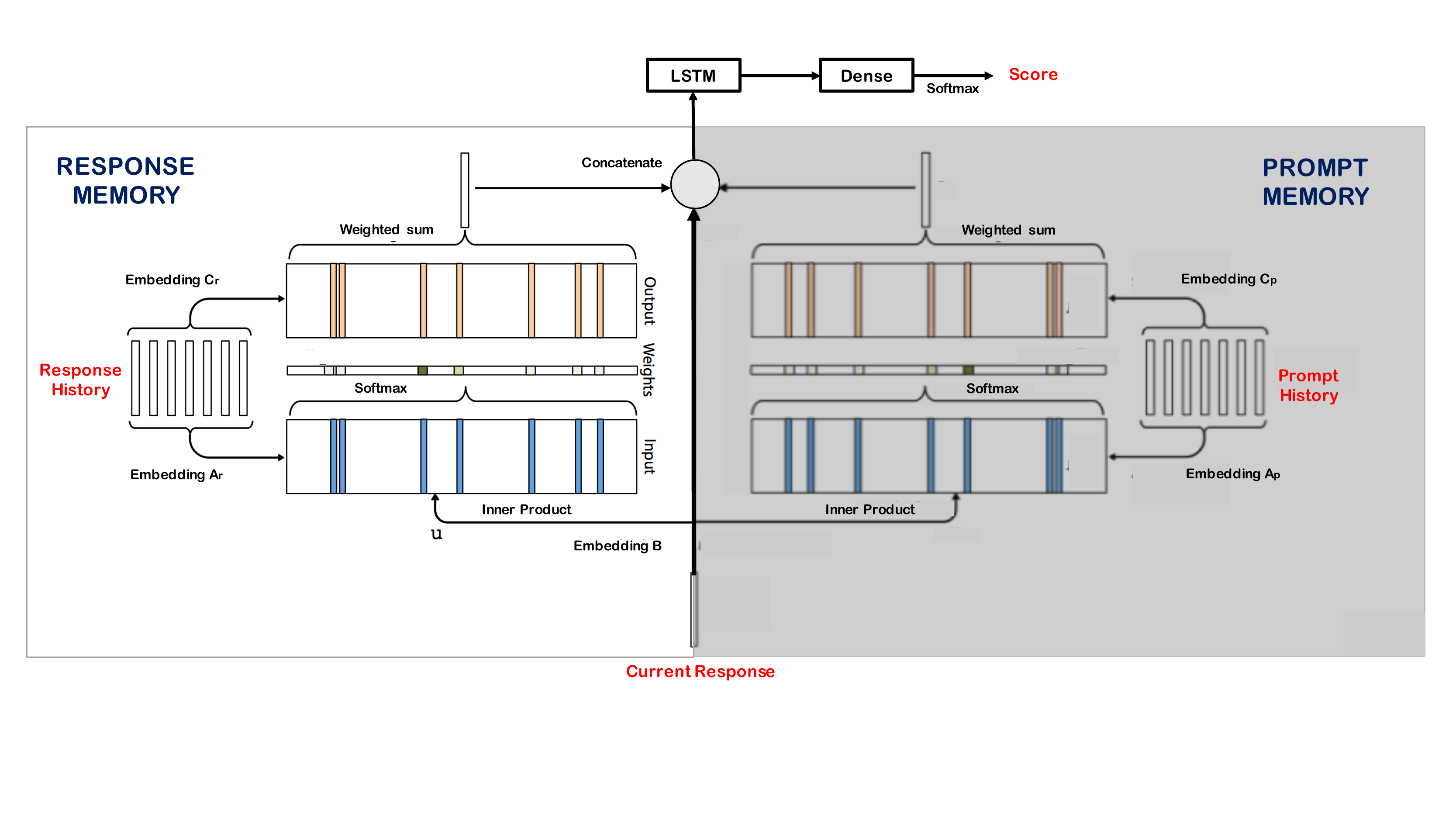}
\vspace{-21.5mm}
\caption{Schematic of a \textit{single hop} module of our modified end-to-end memory network (MemN2N) adapted from \newcite{sukhbaatar15}. Stacking modules on top of each other allows us to model multiple hops.}
%\caption{Score distributions averaged over both raters.}
%\vspace{-2.5mm}
\label{fig:MemN2N}       % Give a unique label
\end{figure*}

\subsection{Recurrent Architectures (BiLSTMs) with and without Attention}

Recurrent architectures, such as Long Short-Term Memory (LSTM) networks, are able to learn long-term dependencies \cite{hochreiter1997long} and are effective in many NLP tasks related to dialog and turn-taking scenarios \cite{ghosh2018sarcasm,skantze2017towards}. We implement the \textit{stacked} BiLSTM network architecture with context attention \cite{yang2016hierarchical}. Here the output of the first BiLSTM hidden layer is fed as input into the subsequent BiLSTM hidden layer. We experimented with varying depths of the stack and empirically selected depth=2. The attention mechanism is as follows. Let the number of words in the dialog $d$ be $w$ and the hidden representation for word $w_{d_{i}}$ be $h_{d_{i}}$. We introduce a word-level attention mechanism where the word representation $u_{d_{i}}$ is weighted by measuring similarity with a word level context vector $u_{d_{w}}$, i.e., randomly initialized and jointly learned during the training. Finally, we compute the dialog vector $v_d$ that summarizes the weighted sum of the word annotations based on the weights. 

\begin{gather}
u_{d_{i}} = \tanh(W_{d}h_{d_{i}} + b_{w})\\
v_{d} = \sum_{i\in[1,w]} \alpha_{{d}_{i}}h_{{d}_{i}}
\end{gather}

where attention $\alpha_{d_{i}}$ is calculated as:
\begin{equation}
\alpha_{{d}_{i}} = \frac{\exp(u_{{d}_{i}}^Tu_{{d}_{w}} )}{\sum_{i\in[1,w]}\exp(u_{{d}_{i}}^Tu_{{d}_{w}} )}
\end{equation}

Figure \ref{figure:bilstm} represents the high-level structure of the BiLSTM Attention architecture. Words are represented as embeddings (e.g., GloVe embedding \cite{pennington14}, the orange blocks in the Figure \ref{figure:bilstm}) and fed to the BiLSTM network. For brevity, we are showing only one BiLSTM layer composed of the \emph{forward} and \emph{backward} layer accounting to the hidden layer $h_{d_{i}}$ (green blocks). Next, context vector $u_{d_{w}}$ is utilized to generate word level attention $\alpha_{{d}_{i}}$ (purple blocks). Finally, the dialog vector $v_d$ passes through a dense+Softmax layer to predict the score of the construct in the given experiment.

%based on space can put the eqns.
%We used the same CV folds used in the feature engineering approach (Section \ref{sec:skll}) in our deep learning experiments as well for the sake of consistency. However, 
To tune the hyperparameters for BiLSTM based experiments, we split the training data for each CV fold into 80\% $train$ and 20\% $dev$, and use the $dev$ partition for parameter tuning. 
We employ the following hyperparameters for the BiLSTM architectures: GloVe embeddings (100D), mini-batch size of 16, recurrent dropout value of 0.3, 10 epochs (with an early-stopping patience of 5), and the Adam optimizer with its default parameters.

\subsection{End to End Memory Networks (MemN2Ns)}

We also investigated the efficacy of the End to End Memory Network (MemN2N) architecture \cite{sukhbaatar15,chen16} adapted to the dialog scoring task, described in \citet{ramanarayanan2019scoring}.  See Figure \ref{fig:MemN2N}. %Similar to recent work \cite{qian19}
The end to end MemN2N architecture models dependencies in text sequences using a recurrent attention model coupled with a memory component, and is therefore suited to modeling how response and prompt histories contribute to a dialog score. We modified the original MemN2N architecture in the following ways: (i) instead of the original \textit{(query, fact history, answer)} tuple that is used to train the network, we have an \textit{(current response, response history, prompt history, answer)} tuple in our case. In other words, we not only embed and learn memory representations between the current response and the history of previous responses, but the history of prior system prompts that have been encountered thus far; (ii) we used an LSTM instead of a matrix multiplication at the final step of the network before prediction;  (iii) we train the network at the turn level, and assign the dialog-level score as the median score of all scores predicted by the network at the turn-level, as mentioned earlier. %For more details about the network architecture and hyperparameter selection, please see \citet{ramanarayanan2019scoring}. 

% Description of the variety of parameters and subsequent experiments (e.g., GloVe vs. word2vec embedding) are reported in the Appendix section.   
We tuned hyperparameters of the network using the hyperas toolkit\footnote{\texttt{http://maxpumperla.com/hyperas/}}. This included the number of neurons in the Dense and LSTM layers as well as the addition of Dropout layers after each memory component. We trained the network for 40 epochs (but with an early-stopping patience of 5, so we generally did not exceed 10 epochs in practice). We experimented with 1, 2 and 3 memory hops and found 2 to be optimal. We found that initializing the memory embedding matrics with pretrained \textit{word2vec} \cite{mikolov13} or \textit{GloVe} \cite{pennington14} embeddings worked better than randomly-initialized ones for prompt history encoding in particular. %, but this did not hold for response history encoding.

%We investigated different embeddings (e.g., GloVe \cite{pennington14}, word+character, etc.), LSTM dimensions ((a) no pre-trained embedding
%depending upon space we can extend/shortened the description. 

%\textcolor{blue}{Matt, please add in}

%Using \textit{Keras} (and \textit{TensorFlow}), we investigated the potential of a "stack" dual-BiLSTM architecture with a various types of embeddings: 1) no pre-trained embeddings, 2) pre-trained embeddings for words alone, and 3) pre-trained embeddings for words and characters. In each of the latter cases, we experimented with 300-dimensional \textit{Google word2vec} \cite{mikolov13} and \textit{GloVe} \cite{pennington14} embeddings separately. For number of LSTM units, word LSTM units, and character LSTM units, we used 100. We also set dropout to 0.25 and used sparse categorical cross-entropy as our loss function. Lastly, we used the \textit{Nadam} optimizer. Training data was first split into 80\% train and 20\% validation and training was run for 8 epochs (with an early-stopping patience of 5).

\subsection{Transformer Models}

The final class of models we explore is the purely attention-based family of transformer models \cite{vaswani2017attention}. Attention, as we have seen, is a mechanism in the neural network that a model can learn to make predictions by selectively attending to a given set of data (and if we are making prediction for one part of a data sample using other parts of the observation about the same sample, this is self-attention). The amount of attention is quantified by learned weights and thus the output is usually formed as a weighted average. The transformer family of models allows one to model sequence data without using recurrent network units by leveraging a special scaled dot product attention mechanism in an encoder-decoder framework, and thus may be particularly suited to modeling our dialog time series data. 

We primarily experimented with BERT (Bidirectional Encoder Representations from Transformers) pre-trained transformer-based language models \cite{devlin2018bert}. We also experimented with two other transformer-based model architectures -- RoBERTa \cite{liu2019roberta} and DistilBERT \cite{sanh2019distilbert} -- and found them to produce results similar to the standard BERT model, so we report only these results for brevity. 
We used the HuggingFace \textit{transformers} library\footnote{\texttt{github.com/huggingface/transformers}} \cite{Wolf2019HuggingFacesTS} to fine-tune a pre-trained model (\emph{bert-base-uncased}) on our training data for each fold of our 10-fold cross-validation setup and report performance averaged across all folds. We use the following hyperparameters: number of epochs = 5, learning rate = 5e-5, and Adam epsilon = 1e-8. 
%and report all these results in the Appendix.
\iffalse
We also investigated the use of general-purpose transformers such as BERT, RoBERTa, and DistilBERT. %add citation
We used HuggingFace's\footnote{\texttt{github.com/huggingface/transformers}} \textit{transformers} library \cite{Wolf2019HuggingFacesTS} to fine-tune pre-trained models (\emph{bert-base-uncased}) to our data. We used the following hyperparameters: number of epochs was set to 5, learning rate was set to 5e-5, weight decay was set to 0.0, and Adam epsilon was set to 1e-8, and \textit{do\_lower\_case} was turned on. Training data was split into 90\% and 10\% train/validation partitions, respectively. Although we tried several models, we only present the best-performing model for ease of presentation, which was based on \textit{"bert-base-uncased"}. However, all transformer-based models performed similarly.
\fi

% \begin{sidewaystable}
\begin{table*}[]
\caption{\label{tab:stats} {\it Automated scoring performance (as measured by the \textbf{quadratic weighted kappa} or QW$\kappa$) of the 6 systems we explore in this paper. Note that we report results for the fusion system with the best QW$\kappa$ (optimized across all combinations of individual systems). The last two columns present Human Inter Rater Agreements for the same data expressed in Krippendorff $\alpha$ and Conger $\kappa$ (note that this is not directly comparable to the reported QW$\kappa$s).}}
\centerline{
\scalebox{0.85}{
\begin{tabular}{  >{\bfseries \em}c | >{\em}l || >{\columncolor[gray]{0.85}}c | >{\columncolor[gray]{0.85}}c | >{\columncolor[gray]{0.85}}c | >{\columncolor[gray]{0.85}}c | >{\columncolor[gray]{0.85}}c | >{\columncolor[gray]{0.85}}c || p{2.1cm} | >{\columncolor[gray]{0.85}}c || >{\columncolor[gray]{0.75} \bfseries}c >{\columncolor[gray]{0.75} \bfseries}c}
\hline
\multirow{2}{*}{Construct}	& \multirow{2}{*}{\textbf{Sub-construct}} &  \textbf{1.} &  \textbf{2.}  &  \textbf{3.}  &  \textbf{4. }  &  \textbf{5. }  &  \textbf{6.} &  \multicolumn{2}{|c|}{\textbf{Fusion Results}}  &  \multicolumn{2}{|c}{\textbf{Human IRR}} \\
	&  &  \textbf{SVM} &  \textbf{SVM++}  &  \textbf{LSTM}   &  \textbf{LSTM$_{att}$}  &  \textbf{MemN2N}  &  \textbf{BERT} &   Best system  & QW$\kappa$ & $\kappa$ & $\alpha$\\
% 	&    & QW$\kappa$     & QW$\kappa$     & QW$\kappa$   & QW$\kappa$   & QW$\kappa$   & QW$\kappa$ & Best system  & QW$\kappa$ & $\kappa$ & $\alpha$ \\
\hline
\multirow{4}{*}{} & Topic &  0.66  & 0.65  & 0.62 & 0.63 & 0.51 & 0.66 & 1, 2, 3, 4, 5, 6 & 0.69 & 0.70 & 0.73\\
\cline{2-2}
Topic & Elaboration	& 0.69 & 0.68 & 0.63 & 0.67 & 0.62 & 0.68 & 1, 3, 4, 5, 6 & 0.72 & 0.76 & 0.75\\
\cline{2-2}
Development & Structure & 0.68 & 0.69 & 0.65 & 0.64 & 0.61 & 0.67 & 1, 5, 6 & 0.72 & 0.75 & 0.75\\
\cline{2-2}
& Task	&  0.71 & 0.72 & 0.65 & 0.69 & 0.62 & 0.71 & 1, 3, 5, 6 & 0.74 & 0.72 & 0.74\\
\hline
\multirow{4}{*}{Interaction} & Engagement & 0.70 & 0.70 & 0.68 & 0.67 & 0.66 & 0.70 & 1, 3, 6 & 0.73 & 0.69 & 0.72\\
\cline{2-2}
& Turn Taking	& 0.67 & 0.68 & 0.63 & 0.63 & 0.62 & 0.70 & 2, 5, 6 & 0.72 & 0.71 & 0.74\\
\cline{2-2}
& Repair & 0.61 & 0.61 & 0.54 & 0.59 & 0.57 & 0.63 & 1, 3, 5, 6 & 0.69 & 0.73 & 0.72 \\
\cline{2-2}
& Appropriateness  & 0.67 & 0.67 & 0.63 & 0.65 & 0.57 & 0.67 & 2, 6 & 0.69 & 0.70 & 0.72\\
\hline
\multicolumn{2}{c|}{\textbf{\emph{Overall Holistic Performance}}}  & 0.72 & 0.71 & 0.66 & 0.69 & 0.64 & 0.71 & 1, 3, 4, 6 & 0.73 & 0.75 & 0.75\\
\hline
\end{tabular}
}
}
\vspace{2mm} %[in some *CL conferences \vspace  = desk reject]
\end{table*}
% \end{sidewaystable}

\section{Observations and Results}

\begin{figure*}[htb]
\begin{subfigure}{\textwidth}
  \centering
  % TOPIC
  \includegraphics[width=.75\linewidth]{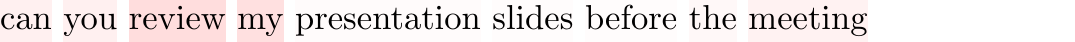}  
  \caption{\emph{Topic}}
  \label{figure:topic_attn}
\end{subfigure}
\begin{subfigure}{\textwidth}
  \centering
  % TASK
  \vspace{2mm}
  \includegraphics[width=.75\linewidth]{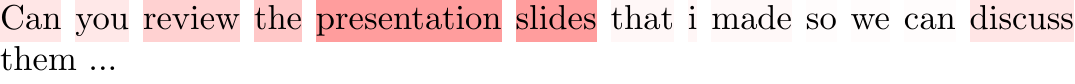}  
  \caption{\emph{Task}}
  \label{figure:task_attn}
\end{subfigure}
\begin{subfigure}{\textwidth}
  \centering
  % APPROPRIATENESS
  \vspace{2mm}
  \includegraphics[width=.75\linewidth]{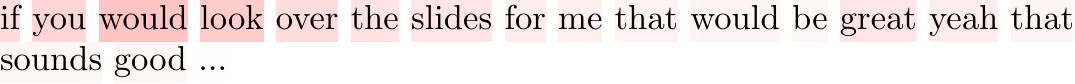}  
  \caption{\emph{Appropriateness}}
  \label{figure:appro_attn}
\end{subfigure}
\begin{subfigure}{\textwidth}
  \centering
  % OVERALL
  \vspace{2mm}
  \includegraphics[width=.75\linewidth]{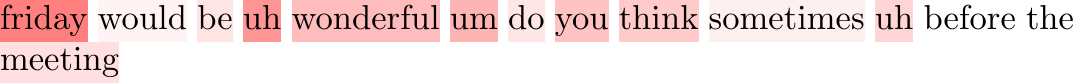}  
  \caption{\emph{Overall holistic score}.}
  \label{figure:overall_attn}
\end{subfigure}
\caption{Attention weights for different scoring constructs obtained from the BiLSTM with attention model.}
\label{fig:attention_weights}
\end{figure*}

% General overview
Table \ref{tab:stats} shows quadratic weighted kappa ($QW\kappa$) values produced by the different automated scoring methods explored in this study. For a more comprehensive report of both accuracy and $QW\kappa$ metrics, refer to Table \ref{tab:stats_full} in the Appendix. Notice that all systems generally produce accuracy numbers in the $0.6 - 0.7$ range, with the BERT and SVM systems (with hand-engineered content features) performing best individually. The final two columns of Table \ref{tab:stats} display two inter-rater agreement statistics -- Conger $\kappa$ and Krippendorff $\alpha$ -- for the human expert scores assigned to the data. Recall that each dialog was scored by 3 out of 8 possible raters. We observe a moderate to high agreement between raters for all dimensions of the scoring rubric, which is not too surprising given that all our raters had significant experience in rating monologic speech data. 

Additionally, it is interesting to note that the $QW\kappa$ of the fusion system is in a similar ballpark as the $\kappa$ and $\alpha$ metrics for human inter-rater agreement across all constructs examined, even slightly higher in some cases such as the task, engagement, and turn-taking constructs. Note however that the $QW\kappa$ values are not directly comparable to the Conger $\kappa$ values, and the human inter-rater agreement values are more of a reference point than a benchmark value. 

We observe that the best fusion systems across constructs all involve the SVM (either with or without politeness features) and BERT systems, suggesting that a combination of feature engineering of content and grammar features along with a neural model leveraging principled attention mechanisms perform best at this automated scoring task. Additionally, we see that MemN2N memory networks make a useful contribution in predicting the constructs of \emph{turn taking}, \emph{repair}, and \emph{topic development}, all constructs that require one to take prior conversational history of the dialog into explicit account in a principled manner. LSTM models (either without or with attention) were part of the best fusion systems for \emph{topic}, \emph{elaboration}, \emph{engagement} and \emph{overall holistic performance}, which require evaluation at the level of the entire dialog. 

As stated earlier, we re-implemented the features from \newcite{ramanarayanan2019scoring} and the SVM experiments are reported as {SVM} in Table \ref{tab:stats} (and Table \ref{tab:stats_full}). The {SVM++} system includes features capturing politeness in the discourse (see Section \ref{table:politeexamples}). Also note that {SVM} experiments and {SVM++} are denoted as systems 1 and 2 respectively for clarity and brevity. We notice that lexicon features capturing politeness help the {SVM++} system achieve better accuracy, particularly for the \emph{structure}, \emph{turntaking}, and \emph{appropriateness} constructs, which is in line with our expections, given that our dialog task requires speakers to use appropriate strategies such as greeting, gratitude, and appreciation, among others, in order to accomplish the task successfully. %strategies capturing different attributes of politeness are able to predict scores of the constructs correctly.       

The BiLSTMs with attention (marked as $LSTM_{attn}$ in Table \ref{tab:stats} or system number 4) perform better compared to the vanilla BiLSTM networks (system number 3) for all the constructs. We positioned an attention layer on top of the stack networks, which means the attention mechanism is able to identify the key characteristics of the constructs. We analyze the heat maps of the attention weights to obtain a better understanding of the model performance. For brevity, we discuss only a couple of constructs here. Each example depicted in the Figure \ref{fig:attention_weights} depicts heat map of the words from a portion of the dialog data corresponding to a request. We chose dialogs which obtained a median human score of 4 (i.e., high proficiency) and were correctly classified by the BiLSTMs with attention model.
% Figure \ref{fig:attention_weights} depicts heat map of the words from a request. 
We observe that words such as ``meeting'' and ``discussion'' receive high weights for the \emph{topic} construct (Figure \ref{figure:topic_attn}). Likewise, Figure \ref{figure:task_attn} also shows that the words representing actions, such as ``reviewing slides'' or ``discussion'' received the highest weights for the \emph{task} construct. For \emph{appropriateness}, we observe words representing positive and respectful tone (e.g., ``if you would look''; ``great yeah'') receiving higher attention weights (Figure \ref{figure:appro_attn}). Finally, in the Figure \ref{figure:overall_attn} we observe the heat map for \emph{overall holistic performance}. Besides key terms such as ``Friday'' (part of the task as well as the automated agent's responses), we observe that positive sentiment words such as ``wonderful'' receive higher attention weights, suggesting that maintaining a positive intonation is weighted more by the BiLSTM with attention model.  

Finally, the results from BERT is reported as System 6 in the Table \ref{tab:stats}. We observe, BERT consistently performs best or comparable to the best model(s) across all the constructs. This verifies the superiority of the transformer architecture, also  observed in prior dialog-act classification task \cite{chakravarty2019dialog}.

% \begin{figure}[t]
% \centering
% \begin{framed}
% \includegraphics[width=7cm]{topic.pdf}
% \caption{Attention weights depicting the \emph{topic} construct}
% \label{figure:topic_attn}
% \end{framed}
% \end{figure}

% \begin{figure}[t]
% \centering
% \begin{framed}
% \includegraphics[width=7cm]{task.pdf}
% \caption{Attention weights depicting the \emph{task} construct}
% \label{figure:task_attn}
% \end{framed}
% \end{figure}

% \begin{figure}[t]
% \centering
% \begin{framed}
% \includegraphics[width=7cm]{appro.pdf}
% \caption{Attention weights depicting the \emph{Appropriateness} construct}
% \label{figure:appro_attn}
% \end{framed}
% \end{figure}

% \begin{figure}[t]
% \centering
% \begin{framed}
% \includegraphics[width=7cm]{overall.pdf}
% \caption{Attention weights depicting the \emph{Overall Score} construct}
% \label{figure:overall_attn}
% \end{framed}
% \end{figure}

% Topic development
% include examples of attention plots here for selected subconstructs.

% Interaction
% include examples of attention plots here for selected subconstructs.

% Overall holistic score

% Fusion results and systems
\section{Implications for Future Work on Automated Dialog Scoring}
\label{sec:discussion}

% Recap
We have presented, for the first time, a comprehensive set of experiments exploring different architectures for machine scoring of text dialog data. We examined both feature-engineered machine learning models as well as multiple neural architectures -- recurrent, memory and attention based -- and found that different optimal combinations of these architectures were useful in scoring different constructs of the text dialog. A combination of carefully selected features along with principled attention-based models were particularly effective. 

% Relative performance of different models: what does this mean and what are the implications
We still need to obtain deeper insights into the relative performance and utility of each system, particularly with respect to interpretability and understandability of the predicted score. Such explainability is particularly crucial in order to deploy such automated scoring systems in the real world. 
Our initial investigation has provided us with some clues, however. Examining the attention weights in particular provides us with some insights regarding the types of word features that the system ``attends'' to. Additionally, we conducted a cursory empirical analysis of errors made by each of our systems. We observed that the errors made by the SVM and BERT systems generally involved an underestimation of the median human score, while the Stack BiLSTM and MemN2N generally overestimated it. We need to conduct further research to unpack these findings fully, and will focus on this as an active line of future work. 

% Future work
This work informs two other important future learning and assessment directions and goals in addition to enhancing the interpretability and explainability of our automated scoring models. The first of these is to leverage insights from understanding these models to develop targeted and personalized feedback to language learners interacting with the dialog system. Indeed it is more crucial to provide learners with actionable feedback (as close to something a human teacher might provide them) than just a numeric score if one has to develop effective and widely-adopted learning applications.  Second, we would like to extend our modeling and analysis experiments of text dialog to speech and spoken dialog.  

% What does this mean, overall, for automated scoring? 
We are poised to observe an increase demand for automated scoring of dialog as conversational systems for language learning and assessment continue to burgeon across the global educational landscape. While this field is still maturing (relative to automated scoring of essays or monologic speech), this paper puts forth a concrete contribution to our understanding of text dialog scoring, paving the path for more comprehensive solutions going forward.

\bibliography{anthology,dialog_scoring_SIGDIAL2020}
\bibliographystyle{acl_natbib}

\appendix
\newpage
\clearpage

% \begin{landscape}
\section{Comprehensive Human and Machine Score Statistics}

\setcounter{table}{0}
\renewcommand{\thetable}{A\arabic{table}}

% \begin{table*}[]
\begin{sidewaystable}[]
\caption{\label{tab:stats_full} {\it Complete table of human and machine score performance statistics reporting both accuracy and quadratic weighted kappas.}}
\centerline{
% \hspace{-10cm}
\scalebox{0.9}{
\begin{tabular}{  >{\bfseries \em}c | >{\em}l || c >{\columncolor[gray]{0.85}}c | c >{\columncolor[gray]{0.85}}c | c >{\columncolor[gray]{0.85}}c | c >{\columncolor[gray]{0.85}}c | c >{\columncolor[gray]{0.85}}c | c >{\columncolor[gray]{0.85}}c || p{2.1cm} | c >{\columncolor[gray]{0.85}}c || >{\columncolor[gray]{0.75} \bfseries}c >{\columncolor[gray]{0.75} \bfseries}c}
\hline
\multirow{2}{*}{Construct}	& \multirow{2}{*}{\textbf{Sub-construct}} &  \multicolumn{2}{|c|}{\textbf{1. SVM}} &  \multicolumn{2}{|c|}{\textbf{2. SVM++}}  &  \multicolumn{2}{|c|}{\textbf{3. Stack BiLSTM}\footnote{We report results for the Stack BiLSTM system with the best QW$\kappa$ (optimized across different types of word and character embeddings).}}   &  \multicolumn{2}{|c|}{\textbf{4. BiLSTM (attn.)}}  &  \multicolumn{2}{|c|}{\textbf{5. MemN2N}}  &  \multicolumn{2}{|c|}{\textbf{6. BERT}} &  \multicolumn{3}{|c|}{\textbf{Fusion Results}\footnote{We report results for the fusion system with the best QW$\kappa$ (optimized across all combinations of individual systems).}}  &  \multicolumn{2}{|c}{\textbf{Human IRR}\footnote{Human Inter Rater Agreements expressed in Krippendorff $\alpha$ and Conger $\kappa$ (note that this is not directly comparable to the reported QW$\kappa$s). }} \\
	&   & ACC & QW$\kappa$    & ACC & QW$\kappa$    & ACC & QW$\kappa$  & ACC & QW$\kappa$  & ACC & QW$\kappa$  & ACC & QW$\kappa$ & Best system & ACC & QW$\kappa$ & $\kappa$ & $\alpha$ \\
\hline
\multirow{4}{*}{} & Topic &  0.71 & 0.66 & 0.70 & 0.65 & 0.68 & 0.62 & 0.70 & 0.63 & 0.68 & 0.51 & 0.70 & 0.66 & 1, 2, 3, 4, 5, 6 & 0.70 & 0.69 & 0.70 & 0.73\\
\cline{2-2}
Topic & Elaboration	& 0.69 & 0.69 & 0.68 & 0.69 & 0.64 & 0.63 & 0.66 & 0.67 & 0.65 & 0.62 & 0.68 & 0.68 & 1, 3, 4, 5, 6 & 0.70 & 0.72 & 0.76 & 0.75\\
\cline{2-2}
Development & Structure & 0.68 & 0.68 & 0.69 & 0.69 & 0.64 & 0.65 & 0.66 & 0.64 & 0.65 & 0.61 & 0.68 & 0.67 & 1, 5, 6 & 0.69 & 0.72 & 0.75 & 0.75\\
\cline{2-2}
& Task	&  0.71 & 0.71 & 0.71 & 0.72 & 0.66 & 0.65 & 0.68 & 0.69 & 0.64 & 0.62 & 0.70 & 0.71 & 1, 3, 5, 6 & 0.71 & 0.74 & 0.72 & 0.74\\
\hline
\multirow{4}{*}{Interaction} & Engagement & 0.70 & 0.70 & 0.70 & 0.70 & 0.66 & 0.68 & 0.67 & 0.67 & 0.66 & 0.66 & 0.68 & 0.70 & 1, 3, 6 & 0.68 & 0.73 & 0.69 & 0.72\\
\cline{2-2}
& Turn Taking	& 0.69 & 0.67 & 0.70 & 0.68 & 0.65 & 0.63 & 0.66 & 0.63 & 0.67 & 0.62 & 0.71 & 0.70 & 2, 5, 6 & 0.70 & 0.72 & 0.71 & 0.74\\
\cline{2-2}
& Repair & 0.66 & 0.61 & 0.66 & 0.61 & 0.60 & 0.54 & 0.63 & 0.59 & 0.64 & 0.57 & 0.65 & 0.63 & 1, 3, 5, 6 & 0.66 & 0.69 & 0.73 & 0.72 \\
\cline{2-2}
& Appropriateness  & 0.67 & 0.67 & 0.68 & 0.67 & 0.61 & 0.63 & 0.65 & 0.65 & 0.62 & 0.57 & 0.66 & 0.67 & 2, 6 & 0.68 & 0.69 & 0.70 & 0.72\\
\hline
\multicolumn{2}{c|}{\textbf{\emph{Overall Holistic Performance}}}  & 0.69 & 0.72 &  0.69  & 0.71 & 0.65 & 0.66 & 0.66 & 0.69 & 0.66 & 0.64 & 0.68 & 0.71 & 1, 3, 4, 6 & 0.70 & 0.73 & 0.75 & 0.75\\
\hline
\end{tabular}
}
}
%\vspace{-1mm} [in some *CL conferences \vspace  = desk reject]
% \end{table*}
\end{sidewaystable}

\end{document}